\documentclass{pasj00}
 \draft
 
\begin{document}
\SetRunningHead{Author(s) in page-head}{Running Head}
\Received{2006/06/25}
\Accepted{2006/12/01}

\title{A CO {\it J}=3--2 Survey of the Galactic Center}

\author{%
Tomoharu \textsc{Oka}\altaffilmark{1}
Makoto \textsc{Nagai}\altaffilmark{1}
Kazuhisa \textsc{Kamegai}\altaffilmark{2}
Kunihiko \textsc{Tanaka}\altaffilmark{2}
and
Nobuyuki \textsc{Kuboi}\altaffilmark{1}
}
\altaffiltext{1}{Research Center for the Early Universe and Department of Physics\\  The University of Tokyo, 7-3-1 Hongo, Bunkyo-ku, Tokyo, Japan 113-0033}
\email{tomo@taurus.phys.s.u-tokyo.ac.jp}
\altaffiltext{2}{Institute of Astronomy, Faculty of Science, The University of Tokyo, 2-21-1 Osawa, Mitaka, Tokyo, Japan 181-0015}


\maketitle

\begin{abstract}
We have surveyed the central molecular zone (CMZ) of our Galaxy in the CO {\it J}=3--2 line with the Atacama Submillimeter-wave Telescope Experiment (ASTE).  Molecular gas in the Galactic center shows high {\it J}=3--2/{\it J}=1--0 intensity ratio ($\sim 0.9$) while gas in the spiral arms in the Galactic disk shows the lower ratio ($\sim 0.5$).  The high-velocity compact cloud CO 0.02--0.02 and the hyperenergetic shell CO 1.27+0.01, as well as gas in the Sgr A region exhibit high {\it J}=3--2/{\it J}=1--0 intensity ratio exceeding 1.5.  We also found a number of small spots of high ratio gas.  Some of these high ratio spots have large velocity widths and some seem to associate with nonthermal threads or filaments.  These could be spots of hot molecular gas shocked by unidentified supernovae which may be abundant in the CMZ.     
\end{abstract}

\section{INTRODUCTION}

The Central Molecular Zone (CMZ) of the Galaxy, a region of radius $\sim 200$ pc, contains large amount of dense molecular gas ($n \geq 10^4$ cm$^{-3}$; e.g., Paglione et al. 1998).   It has also been observed that gas temperature is uniformly high in the CMZ ($T_{\rm k}=$30--60 K; Morris et al. 1983).  Molecular gas in the CMZ shows highly complex distribution and kinematics as well as remarkable variety of peculiar features (e.g., Bally et al. 1987; Burton \&\ Liszt 1992).  Its small-scale structure is characterized by arcs/shells and filaments (Oka et al. 1998).  A population of high-velocity compact clouds (HVCCs) is also unique in the CMZ (Oka et al. 1998, 1999, 2001b).  

The CMZ also contains a large amount of hot ($\sim 10^8$ K) plasma (Koyama et al. 1989), which origin has been suggested to be about $10^3$ supernova explosions in the last $10^4$ years (Yamauchi et al. 1990).  The detection of $^{26}$Al $\gamma$-ray line supports the occurrence of such energetic explosions (von Ballmoos, Diehl, \&\ Sch\"uonfelder 1987).  The widespread SiO emission there has been understood as a result of large-scale shocks such as cloud-cloud collisions and/or fossil superbubbles (Mart\'ein-Pintado et al. 1997; H\"uttemeister et al. 1998).  These suggest that the boisterous molecular gas kinematics in the CMZ may be a result of violent release of kinetic energy by a number of supernova explosions.  

To assess the effect of supernova explosions on the kinematics and physical conditions of the CMZ, it is crucial to pick up supernova/molecular cloud interacting zone completely.  The rotational transition lines of carbon monoxide (CO) in submillimeter wavelength are well established, excellent tracers of the supernova/molecular cloud interaction (White 1994; Arikawa et al. 1999; Moriguchi et al. 2005).  This paper presents a brief report of the ongoing large-scale, high-resolution CO {\it J}=3--2 survey of the CMZ.  The full presentation of the data with detailed analyses will be published in the forthcoming paper.  Detailed analyses of physical conditions and results of follow-up observations for several features of great interest are presented in separate papers (Nagai et al. 2006; Tanaka et al. 2006).

\section{OBSERVATIONS}

The CO {\it J}=3--2 (345.79599 GHz) mapping observations were carried out with the Japanese 10 m submillimeter-wave telescope at Pampa la Bola, Chile (ASTE; Kohno et al. 2004).  The observations were done 2005 July 19--25 in good, stable weather conditions.  The telescope has a beam efficiency of 0.6 and FWHM beam size of 22\arcsec\ at 345 GHz.  The pointing of the telescope was checked and corrected every two hours by observing Jupitar and Uranus, and its accuracy was maintained within 2\arcsec (rms).  

The telescope is equipped with a 345 GHz SIS receiver SC345.  The reciever had an IF frequency of 6.0 GHz, and the local oscillator was centered at 339 GHz.  Calibration of the antenna temperature was accomplished by chopping between an ambient temperature load and the sky.  Since the mixer in SC345 works in a double sideband (DSB) mode, we scaled the DSB antenna temperature ($T_{\rm A}^*$) by multiplying $(2/0.6)=3.33$ to get the SSB main-beam temperature ($T_{\rm MB}$).  Typical system noise temperatures during the observations were 150--300 K (DSB) including atmospheric loss.  Absolute scale and reproducibility of intensities were checked by monitoring NGC 6334I several times a day.  We had $T_{\rm MB}({\rm ASTE})=51.2\pm 3.0$ K at the peak, which is consistent with the value in the previous literature [$T_{\rm MB}({\rm CSO})=49$ K; Kraemer \&\ Jackson 1999].  To minimize intensity fluctuation caused by pointing errors,we employed the absorption dip at $V_{\rm LSR}=-6.1$ km s$^{-1}$ for the reproducibility check.  The intensity scale was found to be stable within 3.8 \% ($1 \sigma$) during the observations.  All spectra were obtained with an XF-type autocorrelation spectrometer.  The spectrometer was operated in the 512 MHz bandwidth (1024 channel) mode, which corresponds to a 445 km s$^{-1}$ velocty coverage and a 0.45 km s$^{-1}$ velocity resolution at 345 GHz.  

We mapped a $\Delta l\times \Delta b = 2\arcdeg\times 0.5\arcdeg$ area in the CO {\it J}=3--2 line, collecting 6990 spectra in total.  A 34\arcsec\ grid spacing was chosen in accordance with that of the NRO 45m CO {\it J}=1--0 survey.  The observations were performed by position-switching to a clean reference position, $(l, b)$=$(0\arcdeg, -2\arcdeg)$.  The on-source integration time was 10 seconds for each position and rms noise was 0.3 K in $T_A^*$.  The data were reduced on the NEWSTAR reduction package.  We subtracted baselines of the spectra by fitting linear lines, or if necessary, the lowest order polynomials that produce straight baselines in emission-free velocity ranges.  About 500 spectra required third-order polynomical fits.

\begin{figure}[h]
\begin{center}
\includegraphics[width=6cm]{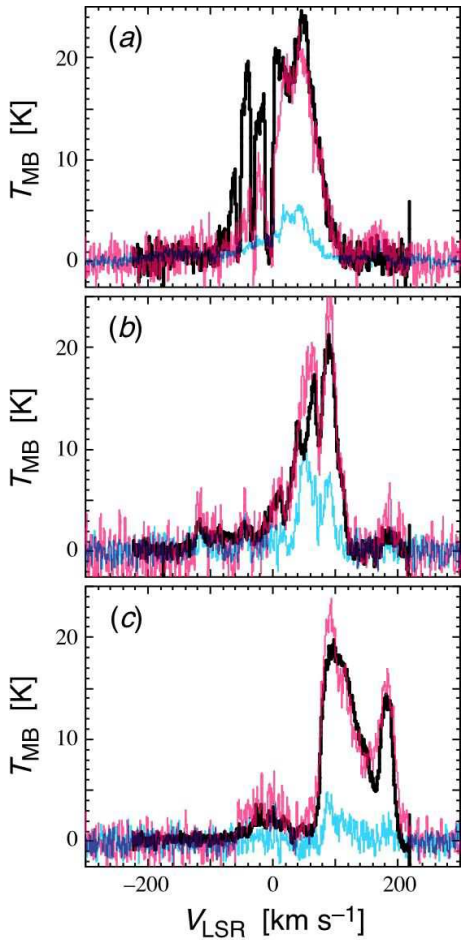}
\end{center}
\caption{Overlay of the CO {\it J}=3--2 (black), CO {\it J}=1--0 (magenta), and $^{13}$CO {\it J}=1--0 (cyan) spectra toward three respective positions, ({\it a}) Sgr A $(l, b)=(-0.05\arcdeg, -0.05\arcdeg)$, ({\it b}) Sgr B2 $(+0.64\arcdeg, -0.06\arcdeg)$, and ({\it c}) $L=1.3\arcdeg$ molecular complex $(+1.28\arcdeg, +0.06\arcdeg)$.  The CO {\it J}=3--2 data are from the ASTE survey, while the both {\it J}=1--0 data sets are from the NRO 45m survey (Oka et al. 1998).  
\label{fig1}}
\end{figure}

\section{RESULTS}

Sample spectra toward three representative positions, Sgr A $(l, b)=(-0.05\arcdeg, -0.05\arcdeg)$, Sgr B2 $(+0.64\arcdeg, -0.06\arcdeg)$, and the center of $L\!=\!1.3\arcdeg$ molecular complex $(+1.28\arcdeg, +0.06\arcdeg)$, are shown in Figure 1.   The CO {\it J}=3--2 data are from the ASTE survey, while the CO {\it J}=1--0 and $^{13}$CO {\it J}=1--0 data are from the NRO 45m telescope (FWHM beam size$=17\arcsec$; Oka et al. 1998).  Generally, each CO {\it J}=3--2 profile in the CMZ is similar in shape and intensity to the CO {\it J}=1--0 profile, suggesting these lines are opaque and thermalized.  In the $^{13}$CO {\it J}=1--0 profiles, the total velocity extent is somewhat smaller than the $^{12}$CO lines, and the expanding molecular ring (EMR; Kaifu 1972; Scoville 1972) is less prominent.  Rather less intense $^{13}$CO {\it J}=1--0 emission from Sgr A and the $L\!=\!1.3\arcdeg$ complex indicates moderate CO opacities.  This is not the case for the opaque Sgr B2 cloud where the $^{13}$CO/$^{12}$CO intensity ratio reaches 0.5.  Toward Sgr A, the {\it J}=3--2 emission is more intense than CO {\it J}=1--0 especially in the negative velocity side.  Such emission with high {\it J}=3--2/{\it J}=1--0 intensity ratio may come from highly excited gas with rather low column density.    

\begin{figure}[h]
\begin{center}
\includegraphics[width=14cm]{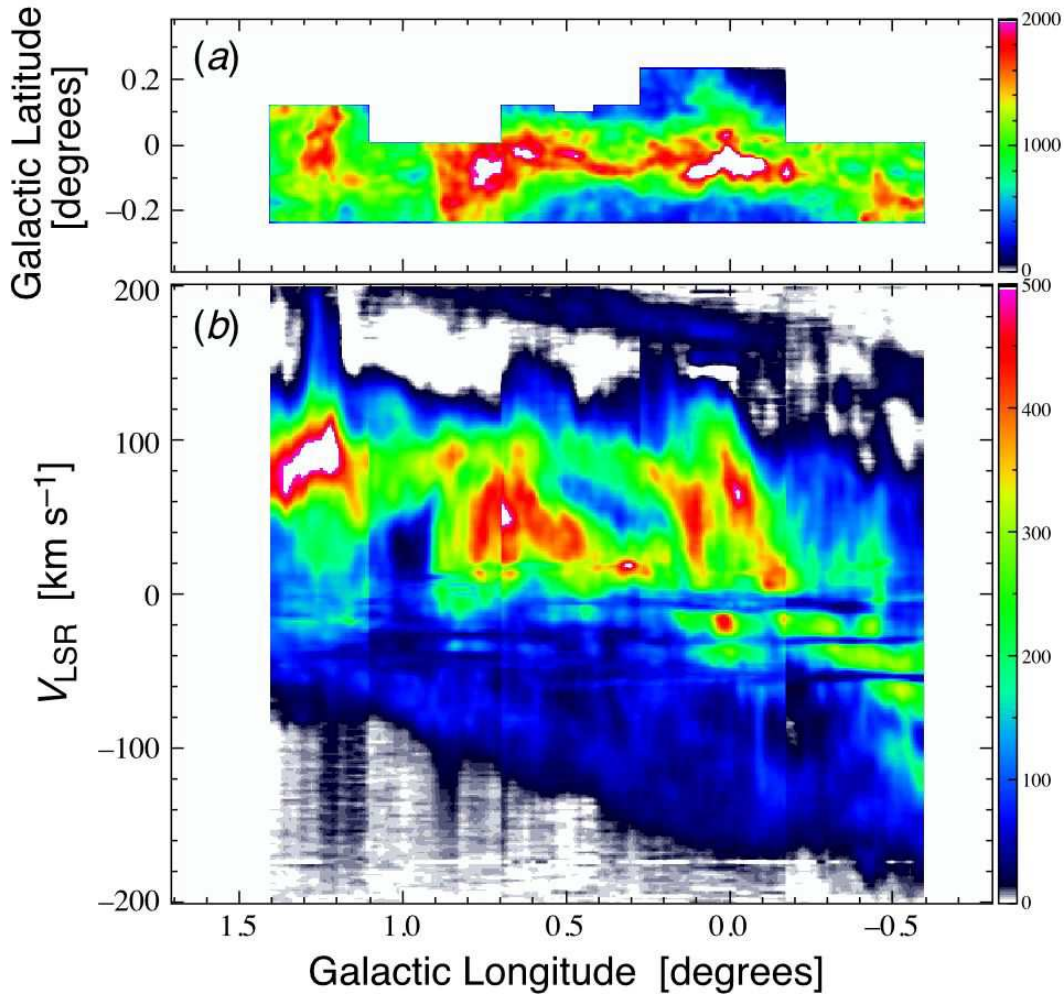}
\end{center}
\caption{({\it a}) Integrated intensity ($\int T_{\rm MB}\,dV$) map of CO {\it J}=3--2 emission.  The emission is integrated over velocities between $V_{\rm LSR}\!=\!-200$ km s$^{-1}$ and $+200$ km s$^{-1}$ and the map has been smoothed to 60\arcsec\ resolution.   ({\it b}) Longitude-velocity map of CO {\it J}=3--2 emission integrated over the observed latitudes ($\sum T_{\rm MB}$).  The map has been smoothed to 60\arcsec\ resolution and summed up to each $+2$ km s$^{-1}$ bin.  
\label{fig2}}
\end{figure}

Figure 2 presents a map of CO {\it J}=3--2 emission integrated over velocities between $V_{\rm LSR}\!=\!-200$ km s$^{-1}$ and $+200$ km s$^{-1}$ ({\it a}), and a longitude-velocity ({\it l-V} map integrated over the observed latitudes ({\it b}).   The CO {\it J}=3--2 data set have been smoothed to 60\arcsec\ spatial resolution and summed up to  each $+2$ km s$^{-1}$ bin.  The CO {\it J}=3--2 emission in the Galactic center extends over the current spatial coverage of the ASTE survey.  Its distribution closely follows characteristics of the CMZ delineated by the CO {\it J}=1--0 surveys (Bally et al. 1987; Burton \&\ Liszt 1992; Oka et al. 1998).  Four major cloud complexes are seen also in the CO {\it J}=3--2 map, from left to right, the $L\!=\!1.3\arcdeg$ complex, the Sgr B complex near $l=0.7\arcdeg$,  the Sgr A complex near $l=0.0\arcdeg$, and the Sgr C complex near $l=0.5\arcdeg$.  The ASTE survey currently does not cover the full spatial extents of the $L=1.3\arcdeg$ and Sgr C complexes.  At the center of the $L=1.3\arcdeg$ complex, we see a clear hole of emission, which is the hyperenergetic shell CO 1.27+0.01 (Oka et al. 2001b).  A wavy filament connects the Sgr B complex to the Sgr A complex.  This appears more prominent in CO {\it J}=3--2 than in CO {\it J}=1--0.  

The longitude-velocity behaviour of CO {\it J}=3--2 emission also resembles that of CO {\it J}=1--0.  Major features in the CMZ can be identified in the {\it l-V} map; including Sgr A, Sgr B, Sgr C, and the $L\!=\!1.3\arcdeg$ complex.   A straight continuous line running from $(l, V_{\rm LSR})=(+1.0\arcdeg, +200\,{\rm km s}^{-1})$ to $(-0.5\arcdeg, +150\,{\rm km s}^{-1})$ is the positive-velocity component of the expanding molecular ring (EMR).   The negative-velocity counterpart of the EMR defines the negative velocity end of CO {\it J}=3--2 emission in the {\it l-V} plane.  Intensity discontinuities in the CO {\it J}=3--2 {\it l-V} map at $l=-0.17\arcdeg$, $+0.28\arcdeg$, $+0.7\arcdeg$, and $+1.11\arcdeg$ are caused by the difference in latitudinal coverage of the survey.  Four spiral arms in the Galactic disk, 3 kpc, 4.5 kpc, local, and +20 km s$^{-1}$ arm in order of velocity, are seen as absorption features.  We see a prominent high-velocity feature at $(l, V_{\rm LSR})=(-0.1\arcdeg, -100\,{\rm km s}^{-1})$.  This is the negative-velocity component of the circumnuclear disk (CND) with its spatial extension.  Another prominent high-velocity feature appears at $(l, V_{\rm LSR})=(+1.25\arcdeg, +160\,{\rm km s}^{-1})$, which is the hyperenergetic shell CO 1.27+0.01.  Although we see faint high-velocity features at $(l, V_{\rm LSR})\simeq (-0.45\arcdeg, -180\,{\rm km s}^{-1})$, $(-0.45\arcdeg, +100\,{\rm km s}^{-1})$, $(+0.2\arcdeg, +140\,{\rm km s}^{-1})$, $(+0.6\arcdeg, +140\,{\rm km s}^{-1})$, and $(+0.85\arcdeg, -120\,{\rm km s}^{-1})$, their entity is controversial.  A faint feature at  $(l, V_{\rm LSR})\simeq (+1.2\arcdeg, -130{\rm km s}^{-1})$ may be a baseline ripple in coarse quality spectra.

\begin{figure}[h]
\begin{center}
\includegraphics[width=7cm]{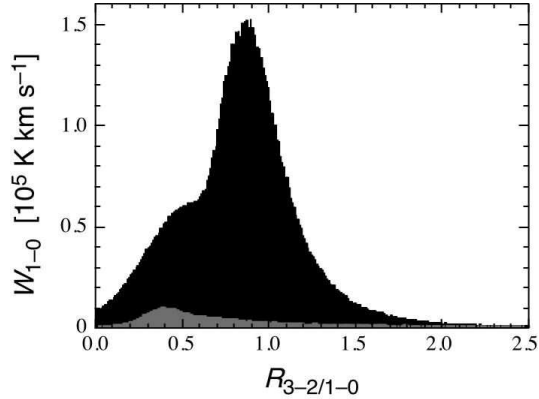}
\end{center}
\caption{Frequency distribution of CO {\it J}=3--2/CO {\it J}=1--0 intensity ratio ($R_{\mbox{3--2/1--0}}$) weighted by the CO {\it J}=1--0 intensity.  Gray area shows the contributions of four spiral arms in the Galactic disk.  
\label{fig3}}
\end{figure}

\section{CO {\it J}=3--2/{\it J}=1--0 Intensity Ratio} 
\subsection{Analyses} 

We often refer ratios between molecular line intensities in diagnoses of physical conditions or chemical compositions of interstellar molecular gas.  Here we use the CO {\it J}=3--2/{\it J}=1--0 intensity ratio ($R_{\mbox{3--2/1--0}}$) to extract highly excited gas from the CMZ, since the ratio is sensitive to the temperature and density of molecular gas.  Figure 3 shows frequency distribution of CO {\it J}=3--2/{\it J}=1--0 intensity ratio weighted by the CO {\it J}=1--0 intensity.  The CO {\it J}=1--0 data are from the NRO 45m survey (Oka et al. 1998).  Both data sets have been smoothed to 60\arcsec\ spatial resolution and summed up to each $+2$ km s$^{-1}$ bin.  The data with 1 $\sigma$ detections in the both lines have been used for the analysis.  The contributions of four foreground spiral arms in the Galactic disk are presented by gray bars.  The spiral arms were defined by straight lines in the {\it l-V} plane, having velocity widths of 2 km s$^{-1}$ (+20 km s$^{-1}$ arm) or 4 km s$^{-1}$ (local, 4.5 kpc, 3 kpc arms).  The $R_{\mbox{3--2/1--0}}$ distribution has a prominent peak at 0.85, which is the typical value in the CMZ, and a shoulder at $\sim 0.5$.  The low-ratio shoulder is mostly attributable to the foreground spiral arms.  The bulk of molecular gas in the CMZ has higher $R_{\mbox{3--2/1--0}}$ than that in the spiral arms in the Galactic disk.  The ratio close to unity indicates that the lowest three rotational levels of CO are thermalized and the transition lines are moderately opaque.

\begin{figure}[h]
\begin{center}
\includegraphics[width=6cm]{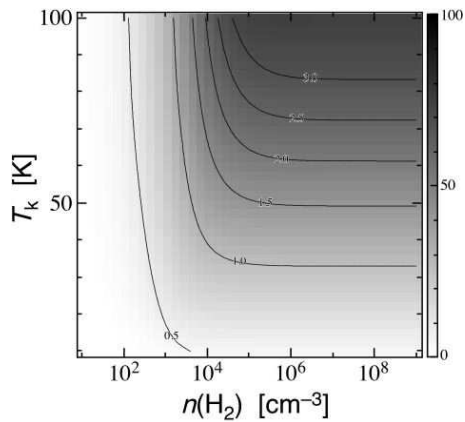}
\end{center}
\caption{Curves of CO {\it J}=3--2/{\it J}=1--0 intensity ratio as a function of hydrogen density [$n({\rm H_2})$] and the kinetic temperature ($T_{\rm k}$) for the CO column density per unit velocity width $N_{\rm CO}/dV=10^{17}$ cm$^{-2}$ (km s$^{-1}$)$^{-1}$.
\label{fig4}}
\end{figure}

Generally, $R_{\mbox{3--2/1--0}}$ can be a measure of temperature and density if the CO column density per unit velocity width is not very high.  High CO {\it J}=3--2/{\it J}=1--0 ratios have been found in UV-irradiated cloud surfaces near early-type stars (e.g., White et al. 1999), and shocked molecular gas adjacent to supernova remnants (e.g., Dubner et al. 2004).  Here we try to extract highly excited gas from the CO data sets by $R_{\mbox{3--2/1--0}}\geq 1.5$.   One-zone LVG calculations say that $R_{\mbox{3--2/1--0}}\geq 1.5$ corresponds to $n({\rm H}_2)\geq 10^{3.6}$ cm$^{-3}$ and $T_{\rm k}\geq 48$ K when $N_{\rm CO}/dV=10^{17}$ cm$^{-2}$ (km s$^{-1}$)$^{-1}$ (Fig.4).

\begin{figure}[h]
\begin{center}
\includegraphics[width=14cm]{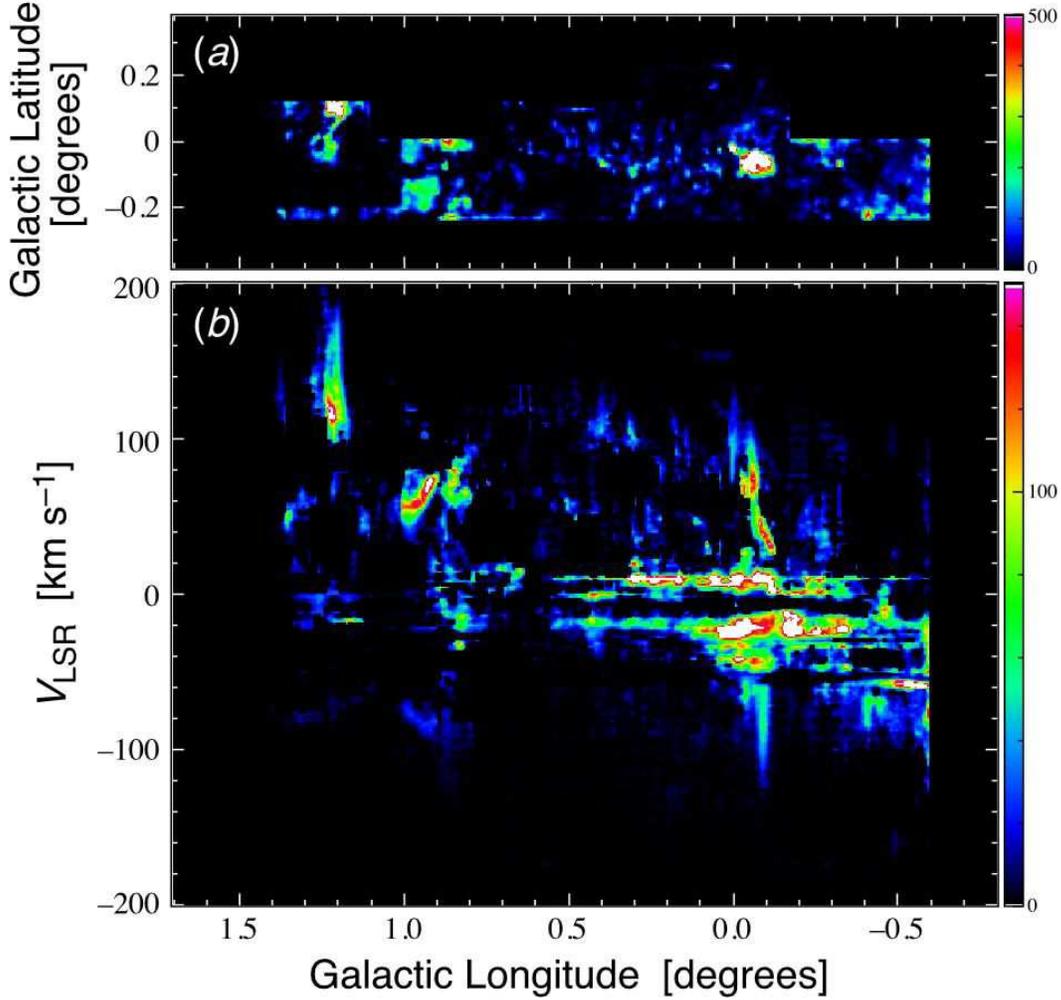}
\end{center}
\caption{({\it a}) A map of CO {\it J}=3--2 emission ($\int T_{\rm MB}\,dV$) integrated over velocities between $V_{\rm LSR}\!=\!-200$ km s$^{-1}$ and $+200$ km s$^{-1}$ for data with $R_{\mbox{3--2/1--0}} \geq 1.5$.  Data severely contaminated by the foreground disk gas, $(-55 + 10\,l)$ km s$^{-1}$ $\leq V_{\rm LSR} \leq +15$ km s$^{-1}$, have been excluded.  ({\it b}) Longitude-velocity map of CO {\it J}=3--2 emission integrated over the observed latitudes ($\sum T_{\rm MB}$) for data with $R_{\mbox{3--2/1--0}} \geq 1.5$ .  
\label{fig5}}
\end{figure}

Figure 5 shows the spatial and longitude-velocity distribution of high $R_{\mbox{3--2/1--0}}$ gas.  Data severely contaminated by the foreground disk gas, $(-55 + 10\,l) \leq V_{\rm LSR} \leq +15$ km s$^{-1}$, have been excluded from the velocity integration in making Fig.5{\it a}.  The spatial distribution of high $R_{\mbox{3--2/1--0}}$ gas in the CMZ shows a number of clumps as well as some diffuse components.  The most prominent one is a clump at Sgr A, which consists of several high-velocity features in both sides of $V_{\rm LSR}=0$ km s$^{-1}$.  Another prominent clump is found in the $L\!=\!1.3\arcdeg$ complex, appearing in the {\it l-V} plane with a extremely broad velocity width.  We also see a number of small spots of high $R_{\mbox{3--2/1--0}}$ gas especially in the inner part of CMZ ($|l|\leq 0.5\arcdeg$).  Some of the high-velocity compact clouds (HVCCs) identified in the {\it J}=1--0 data exhibit high $R_{\mbox{3--2/1--0}}$, although all of them do not necessarily have high ratio.  Two diffuse high $R_{\mbox{3--2/1--0}}$ components are found in the $0.2\arcdeg\!\times\!0.2\arcdeg$ region centered at $l\simeq 0.9\arcdeg$ (the {\it $L\!=\!0.9\arcdeg$ Anomaly}) and in the periphery of the Sgr C complex.  In the velocity range of $-60 \leq V_{\rm LSR} \leq +15$ km s$^{-1}$, we see several narrow-velocity-width features elongated in longitude.    We discuss these high $R_{\mbox{3--2/1--0}}$ features briefly in the following sections.

\begin{figure}[h]
\begin{center}
\includegraphics[width=7cm]{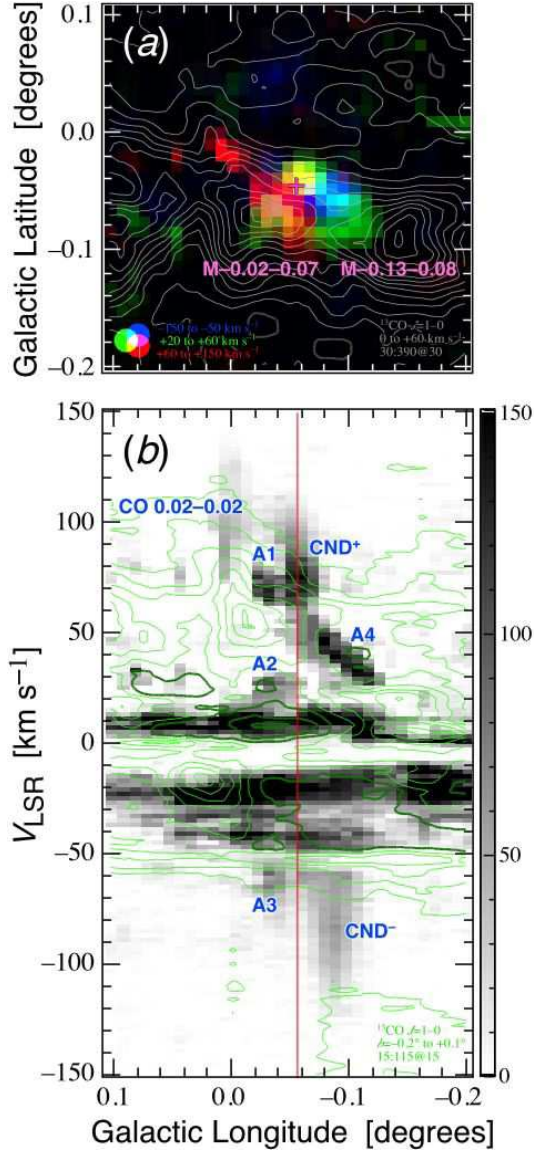}
\end{center}
\caption{({\it a}) A map of CO {\it J}=3--2 emission with $R_{\mbox{3--2/1--0}} \geq 1.5$ in the vicinity of Sgr A.   Integrated intensities for three velocity ranges were presented by different colors; $-150\leq V_{\rm LSR}\leq -50$ km s$^{-1}$ (blue), $+20\leq V_{\rm LSR} \leq +60$ km s$^{-1}$ (green), and $+60\leq V_{\rm LSR} \leq +150$ km s$^{-1}$ (red).  Cross denotes the position of Sgr A$^*$.  Gray contours show a $^{13}$CO {\it J}=1--0 map integrated over velocities $0\leq V_{\rm LSR} \leq +60$ km s$^{-1}$ (red).  Contour interval is 30 K km s$^{-1}$.   ({\it b}) Longitude-velocity map of CO {\it J}=3--2 emission for data with $R_{\mbox{3--2/1--0}} \geq 1.5$.  Latitudinal integration range was from $b=-0.2\arcdeg$ to $+0.1\arcdeg$.  Green contours show an {\it l-V} map of $^{13}$CO {\it J}=1--0 emission integrated over the same latitudes.  A red vertical line denote the longitude of Sgr A$^*$.  Dark thick contours indicate hollows. 
\label{fig6}}
\end{figure}

\subsection{Highly Excited Gas in the CMZ}
\subsubsection{Sgr A Region}

Many authors have discussed the relationship between the radio continuum source Sgr A and the molecular features at scales less than a few arcminutes (e.g., Brown \&\ Lizst 1984: Genzel \&\ Townes 1987).  Our CO {\it J}=3--2 data with 60\arcsec\ resolution shows a high $R_{\mbox{3--2/1--0}}$ clump with the size of $6\arcmin\!\times\!4\arcmin$ at Sgr A and a small redshifted clump in the northeast (Fig. 6{\it a}).  The small red-shifted clump is a well-known HVCC, CO 0.02--0.02 (Oka et al. 1999).   The Sgr A clump splits up several components in the {\it l-V} plane (Fig. 6{\it b}).  We see a pair of high-velocity emission (named CND$^+$ and CND$^-$).  This feature is much larger than the circumnuclear disk (CND; diameter $\sim 100\arcsec$; e.g., Christopher et al. 2005), which is the well-known molecular feature surrounding Sgr A$^*$, and its middle point is about 1\arcmin\ displaced from Sgr A$^*$.  We suggest that CND$^{+-}$ may be a rotating disk-like structure of highly-excited gas, which includes the `classical' CND.  

The component A2 may be a shocked gas clump generated by the interaction with the Sgr A East SNR (Tsuboi et al. 2006).  The components A1 and A3 reside in the same direction as A2, suggesting that they could be relevant to Sgr A East as well.  The component A4 appears in the high-velocity end of the bridge which connects M--0.02--0.07 and M--0.13--0.08, which are well-known dense molecular clouds near Sgr A.  Spatially it follows the Galactic northwestern periphery of M--0.02--0.07.  Its relatively narrow velocity width ($\lesssim 15$ km s$^{-1}$) suggests a non-shock origin.  We speculate that A4 could be a photon-dominated region (PDR) formed by intense UV radiation from the central cluster.

\subsubsection{L=1.3\arcdeg\ Complex}

The $L\!=\!1.3\arcdeg$ complex is the large molecular feature having a prominent elongation toward positive latitude (Oka et al. 1998).  Two expanding shells have been identified at the center of $L\!=\!1.3\arcdeg$ complex (CO 1.27+0.01; Oka et al. 2001b).  The CO {\it J}=3--2 data demonstrate the entity of these expanding shells (Fig.7).  High $R_{\mbox{3--2/1--0}}$ gas is abundant in the Galactic northwestern rim of the `minor' shell.   The simple kinematics of high $R_{\mbox{3--2/1--0}}$ gas associated with the `major' shell can be accounted by a single explosive event, while the `minor' shell has a complex kinematics having a steep velocity gradient in its Galacitc northern part.  High $R_{\mbox{3--2/1--0}}$ gas in the velocities from $+130$ to $+160$ km s$^{-1}$ shows a striking symmetry, which origin is uncertain, however.  The association of high $R_{\mbox{3--2/1--0}}$ gas with these expanding shells ensures that they are generated by a series of supernova explosions, suggesting that a microburst of star formation has occurred there in the recent past.

\begin{figure}[h]
\begin{center}
\includegraphics[width=7cm]{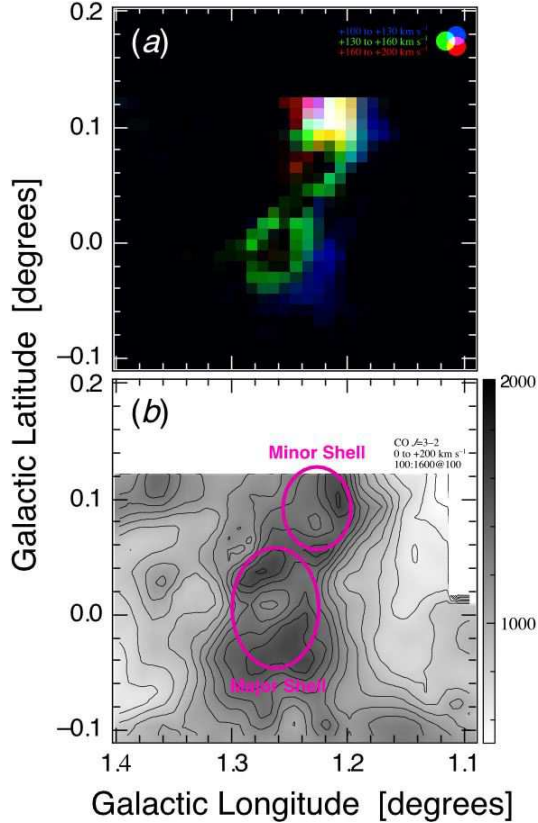}
\end{center}
\caption{({\it a}) A map of CO {\it J}=3--2 emission with $R_{\mbox{3--2/1--0}} \geq 1.5$ in the central region of the $L=1.3\arcdeg$ complex.   Integrated intensities for three velocity ranges were presented by different colors; $+100\leq V_{\rm LSR}\leq +130$ km s$^{-1}$ (blue), $+130\leq V_{\rm LSR} \leq +160$ km s$^{-1}$ (green), and $+160\leq V_{\rm LSR} \leq +200$ km s$^{-1}$ (red).   ({\it b}) A map of CO {\it J}=3--2 emission integrated over velocities from 0 to 200 km s$^{-1}$.  Contour interval is 100 K km s$^{-1}$.  Magenta ellipses denote two expanding shells identified in CO {\it J}=1--0 data (Oka et al. 2001b).  
\label{fig7}}
\end{figure}

\subsubsection{L=0.9\arcdeg\ Anomaly}

We found an anomalous gathering of high $R_{\mbox{3--2/1--0}}$ gas in the $\sim 0.2\arcdeg\times 0.2\arcdeg$ area centerd at $(l, b)\simeq (+0.9\arcdeg, -0.1\arcdeg)$ (Fig.8).  This region corresponds to the edge of the Sgr B complex and the tangent of the 120 pc star forming ring (Arm I; Sofue 1995).  The spatial distribution and kinematics of high $R_{\mbox{3--2/1--0}}$ gas is highly complex.  Roughly, high $R_{\mbox{3--2/1--0}}$ gas in this anomaly can be divided into three classes; (1) spatially extended, red-shifted gas spreading over the $\sim 0.2\arcdeg\times 0.2\arcdeg$ area, (2) compact spots lying in the rim the Sgr B complex, and (3) highly blue-shifted gas at $(l, b, V_{\rm LSR})\simeq (+0.85\arcdeg, 0.0\arcdeg, -120\,{\rm km s}^{-1})$.  

The first class consists of several components, following the lower-velocity periphery of the prominent emission feature at $V_{\rm LSR}\sim +80$ km s$^{-1}$, which bridges the $L\!=\!1.3\arcdeg$ complex and the Sofue's Arm II (Sofue 1995).  The origin of this class of gas is really unknown.  The second class consists of three HVCCs with high $R_{\mbox{3--2/1--0}}$ in the rim of the Sgr B complex; CO 0.86--0.23, CO 0.88--0.08, and CO 0.88+0.00 (see \S 4.2.4).  Their large velocity-widths and high $R_{\mbox{3--2/1--0}}$ suggest that they had been accelerated by supernova blast waves.  Indeed, CO 0.88+0.00 seems to be associated with the TeV $\gamma$-ray bright SNR, G0.9+0.1 (Aharonian et al. 2005).  The third class of gas is also adjacent to the radio shell SNR 0.9+0.1.  The spatial extent of the blue-shifted gas seems to be larger than that of the SNR radio shell.  Although the interaction with G0.9+0.1 or the X-ray pulsar inside is a likely origin of the blue-shifted high $R_{\mbox{3--2/1--0}}$ gas, the way of interaction is quite unknown.

\begin{figure}[h]
\begin{center}
\includegraphics[width=7cm]{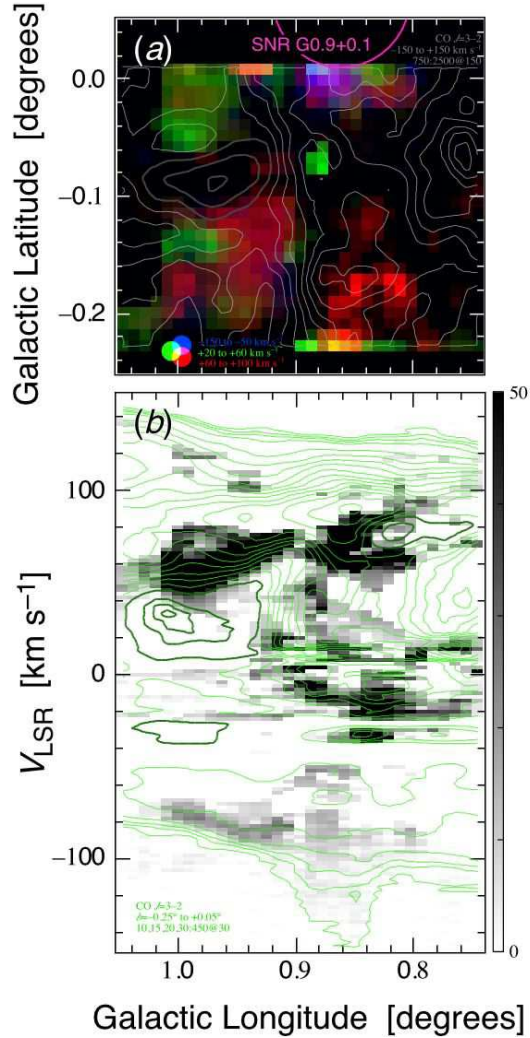}
\end{center}
\caption{({\it a}) A map of CO {\it J}=3--2 emission with $R_{\mbox{3--2/1--0}} \geq 1.5$ of the {\it $L\!=\!0.9\arcdeg$ Anomaly}.   Integrated intensities for three velocity ranges were presented by different colors; $-150\leq V_{\rm LSR}\leq -50$ km s$^{-1}$ (blue), $+20\leq V_{\rm LSR} \leq +60$ km s$^{-1}$ (green), and $+60\leq V_{\rm LSR} \leq +100$ km s$^{-1}$ (red).   Gray contours show a map of CO {\it J}=3--2 emission integrated over velocities $-150\leq V_{\rm LSR} \leq +150$ km s$^{-1}$.  A magenta arc denotes the radio shell of SNR 0.9+0.1 (LaRosa et al. 2000).  ({\it b}) Longitude-velocity map of CO {\it J}=3--2 emission for data with $R_{\mbox{3--2/1--0}} \geq 1.5$ (gray).  Latitudinal integration range was from $b=-0.23\arcdeg$ to $+0.01\arcdeg$.  Green contours show an {\it l-V} map of CO {\it J}=3--2 emission integrated over the same latitudes.  
\label{fig8}}
\end{figure}

\subsubsection{High $R_{\mbox{3--2/1--0}}$ Spots}

We found a number of high $R_{\mbox{3--2/1--0}}$ spots (Fig.9).  Here we present a list of high $R_{\mbox{3--2/1--0}}$ spots (Table 1) for further investigations.  Most of these spots have compact appearances and prefer high-velocity ends of giant molecular clouds within the CMZ.  Many high $R_{\mbox{3--2/1--0}}$ spots have compact entities with large velocity width, exhibiting signs of shocked gas.  Two of those, CO 0.07+0.17 and CO --0.04+0.05, associate with nonthermal `threads'.  CO 0.40--0.07 also overlaps with a faint radio filament.  Four high ratio spots seem to be relevant to the bundle of nonthermal filaments of the Galactic Center Radio Arc.  These facts strongly suggest that supernova/molecular cloud interaction plays an important role in accelerating electrons and forming nonthermal threads and filaments, which are unique and abundant in the CMZ.  It has been reported that shocked molecular gas is associated with the nonthermal filament `Snake' at the intersection with the SNR G359.1--0.5 (Yusef-Zadeh, Uchida, \&\ Roberts 1995; Lazendic et al. 2002).   The association of shocked gas with nonthermal flaments prefers the hypothesis that localized magnetic tubes with a milligauss field are illuminated by relativistic electrons at these filaments.

\begin{figure}[h]
\begin{center}
\includegraphics[width=14cm]{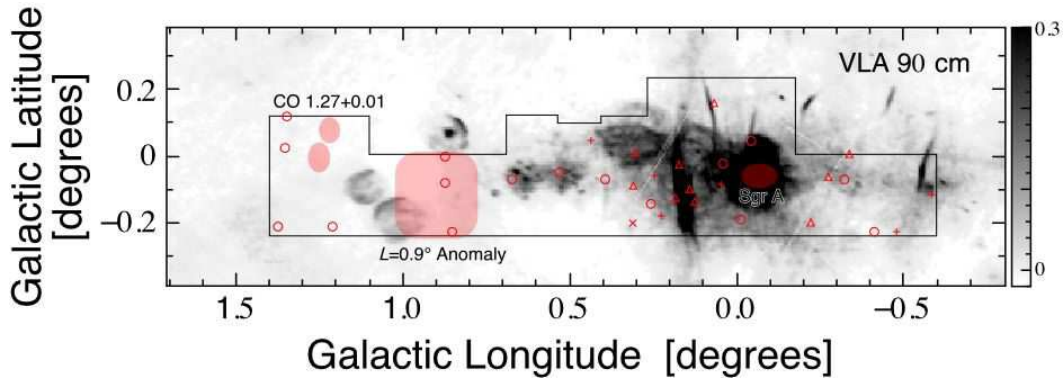}
\end{center}
\caption{Spatial distribution of high ratio ($R_{\mbox{3--2/1--0}}\geq 1.5$) spots superposed on the VLA 90 cm image (LaRosa et al. 2000).  Red circles indicate high-velocity compact clouds (HVCCs) or high-velocity wings with high ratio.   Triangles are high ratio spots in velocity ends of clouds.  Crosses are those in cloud edges.  The `x' denotes a high ratio cloud in the 20 km s$^{-1}$ arm.  
\label{fig9}}
\end{figure}

\begin{table*}
\small
\begin{center}
Table~1.\hspace{4pt}High $R_{\mbox{3--2/1--0}}$ Spots in the CMZ.  \\
\end{center}
\vspace{6pt}
\begin{tabular*}{\columnwidth}{@{\hspace{\tabcolsep}
\extracolsep{\fill}}ccccc}
\hline\hline\\[-6pt]
Name & $l$ & $b$ & $V_{\rm LSR}$  & Comments  \\
 & (deg) & (deg) & (km s$^{-1}$)  &   \\
[4pt]\hline\\[-6pt]
CO --0.59--0.11 \dotfill & $-0.59$ & $-0.11$ & $+55$ to $+95$ & cloud edge? \\
CO --0.48--0.23 \dotfill & $-0.48$ & $-0.23$ & $+90$ to $+110$ & cloud edge \\
CO --0.41--0.23 \dotfill & $-0.41$ & $-0.23$ & $-100$ to $-50$ & HVCC$^{1}$ \\
CO --0.34+0.01 \dotfill & $-0.34$ & $+0.01$ & $-110$ to $-85$ & velocity end \\
CO --0.32--0.07 \dotfill & $-0.32$ & $-0.07$ & $+30$ to $+110$ & HVCC \\
CO --0.27--0.06 \dotfill & $-0.27$ & $-0.06$ & $+30$ to $+45$ & velocity end \\
CO --0.22--0.19 \dotfill & $-0.22$ & $-0.19$ & $+40$ to $+70$ & velocity end of a small cloud \\
CO --0.04+0.05 \dotfill & $-0.04$ & $+0.05$ & $-100$ to $-60$ & high-velocity wing? \\
CO 0.00--0.19 \dotfill & $0.0$ & $-0.19$ & $+70$ to $+110$ & HVCC \\
CO 0.05--0.09 \dotfill & $+0.05$ & $-0.09$ & $+20$ to $+50$ & cloud edge? \\
CO 0.05--0.02 \dotfill & $+0.05$ & $-0.02$ & $+60$ to $+95$ & root of a high-velocity wing \\
CO 0.07+0.17 \dotfill & $+0.07$ & $+0.17$ & $+75$ to $+95$ & velocity end \\
CO 0.12--0.13 \dotfill & $+0.12$ & $-0.13$ & $+40$ to $+55$ & velocity end of CO 0.13--0.13$^{2}$ \\
CO 0.14--0.09 \dotfill & $+0.14$ & $-0.09$ & $+70$ to $+110$ & velocity end \\
CO 0.17--0.02 \dotfill & $+0.17$ & $-0.02$ & $+70$ to $+95$ & velocity end \\
CO 0.18--0.12 \dotfill & $+0.18$ & $-0.12$ & $+95$ to $+120$ & velocity end \\
CO 0.23--0.18 \dotfill & $+0.22$ & $-0.18$ & $+50$ to $+100$ & edge of the expanding cavity$^{3}$ \\
CO 0.25--0.06 \dotfill & $+0.25$ & $-0.06$ & $+35$ to $+60$ & inner rim of the expanding cavity \\
CO 0.26--0.14 \dotfill & $+0.26$ & $-0.14$ & $+40$ to $+70$ & HVCC in the edge of the expanding cavity \\
CO 0.30+0.02 \dotfill & $+0.30$ & $+0.02$ & $+20$ to $+55$ & velocity end \\
CO 0.31--0.09 \dotfill & $+0.31$ & $-0.09$ & $+100$ to $+120$ & velocity end \\
CO 0.31--0.20 \dotfill & $+0.31$ & $-0.20$ & $+20$ & a clump in the 20 km s$^{-1}$ arm \\
CO 0.40--0.07 \dotfill & $+0.40$ & $-0.07$ & $+90$ to $+140$ & high-velocity wing \\
CO 0.43+0.05 \dotfill & $+0.43$ & $+0.05$ & $+80$ to $+120$ & cloud edge? \\
CO 0.54--0.05 \dotfill & $+0.54$ & $-0.05$ & $+110$ to $+140$ & high-velocity wing \\
CO 0.68--0.07 \dotfill & $+0.68$ & $-0.07$ & $+110$ to $+140$ & high-velocity wing \\
CO 0.86--0.23 \dotfill & $+0.86$ & $-0.23$ & $+60$ to $+100$ & HVCC in the L=0.9\arcdeg\ anomaly \\
CO 0.88+0.00 \dotfill & $+0.88$ & $+0.00$ & $-100$ to $-70$ & HVCC associated with SNR 0.9+0.1 \\
CO 0.88--0.08 \dotfill & $+0.88$ & $-0.08$ & $+20$ to $+60$ & HVCC in the L=0.9\arcdeg\ anomaly \\
CO 1.22--0.21 \dotfill & $+1.22$ & $-0.21$ & $+60$ to $+180$ & HVCC \\
CO 1.35+0.12 \dotfill & $+1.35$ & $+0.12$ & $+25$ to $+70$ & high-velocity wing \\
CO 1.36+0.03 \dotfill & $+1.36$ & $+0.03$ & $+45$ to $+70$ & high-velocity wing \\
CO 1.38--0.21 \dotfill & $+1.38$ & $-0.21$ & $+100$ to $+165$ & HVCC \\
\\[4pt]
\hline
\end{tabular*}
\vspace{6pt}
\noindent
$^{1}$ High-velocity compact cloud  \\
$^{3}$ The cloud interacting with the nonthermal filaments of the Radio Arc (Oka et al. 2001a).  \\
$^{3}$ The cavity associated with the nonthermal filaments of the Radio Arc (Oka et al. 2001a).  \\
\end{table*}

\subsection{Absorption by Foreground Gas}

We see several narrow velocity width features in $|l|\leq 0.6$, $-60\leq V_{\rm LSR} \leq +15$ km s$^{-1}$.  They are elongated in longitude, being slightly displaced from the well-known foreground arms in velocity.  It is not likely that they are highly excited gas in the interam region where low density gas dominates.  

Here we made a model with a warm opaque cloud veiled by a layer of cool absorber.  Parameters were chosen $n({\rm H_2})=10^4$ cm$^{-3}$, $T_{\rm k}=50$ K, $N_{\rm CO}/dV=10^{18}$ cm$^{-2}$ (km s$^{-1}$)$^{-1}$ for the warm cloud, and $n({\rm H_2})=10^2$ cm$^{-3}$, $T_{\rm k}=10$ K for the cool absorber.  Figure 10 shows the results of LVG calculations, CO line intensities as functions of $N_{\rm CO}/dV$ of the cool absorber.  Since the {\it J}=2 level is subthermally excited in the cool absorber, the {\it J}=3--2 line is hardly absorbed unless $N_{\rm CO}/dV\geq 10^{16}$ cm$^{-2}$ (km s$^{-1}$)$^{-1}$ where radiative excitation by the photon-trapping process dominates.  This effect raises the {\it J}=3--2/{\it J}=1--0 intensity ratio without participation of highly excited, less opaque gas.  

\begin{figure}[h]
\begin{center}
\includegraphics[width=8cm]{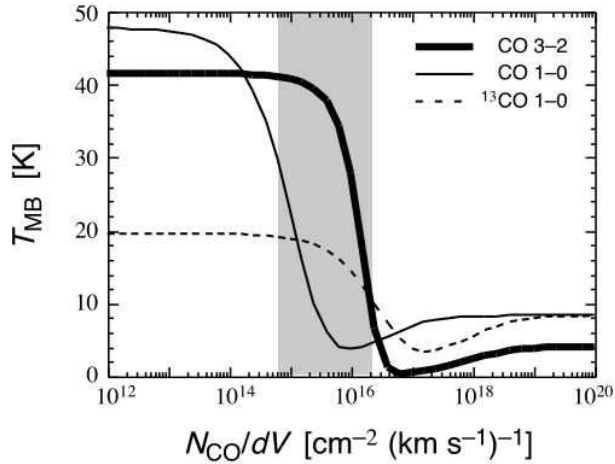}
\end{center}
\caption{Curves of CO {\it J}=3--2 (thick solid line), CO {\it J}=1--0 (thin solid line), and $^{13}$CO {\it J}=1--0 (broken line) antenna temperatures as functions of the CO column density per unit velocity width $N_{\rm CO}/dV$ of the cool absorber.  Shaded area indicates the range of $N_{\rm CO}/dV$ where the {\it J}=3--2/{\it J}=1--0 ratio exceeds 1.5.  
\label{fig10}}
\end{figure}

\section{SUMMARY}

This paper presents a brief report of ongoing large-scale CO {\it J}=3--2 survey of the Galactic center with Atacama Submillimeter-wave Telescope Experiment (ASTE).  We have mapped a $\Delta l \times \Delta b = 2\arcdeg\times 0.5\arcdeg$ region of the CMZ.  

The CO {\it J}=3--2 distribution closely follows characteristics of the CMZ delineated by the CO {\it J}=1--0 surveys, while it shows several faint high-velocity features previously undetected.  The analysis of the {\it J}=3--2/{\it J}=1--0 intensity ratio shows that the bulk of molecular gas in the CMZ is thermalized and moderately opaque.  We found clumps of high $R_{\mbox{3--2/1--0}}$ gas at the Sgr A region, the high-velocity compact cloud CO 0.02--0.02, and the hyperenergetic shell CO 1.27+0.01.  We also found a number of small spots of $R_{\mbox{3--2/1--0}}$ gas in the inner part of CMZ.  Some of the high $R_{\mbox{3--2/1--0}}$ spots have entities with large velocity width and some seem to associate with nonthermal threads or filaments.  Two diffuse high $R_{\mbox{3--2/1--0}}$ components were found in the $0.2\arcdeg\!\times\!0.2\arcdeg$ region centered at $l\simeq 0.9\arcdeg$ ({\it $L=0.9\arcdeg$ Anomaly}) and in the periphery of the Sgr C complex.  

Most of these high $R_{\mbox{3--2/1--0}}$ features are likely shocked gas possibly generated by the interaction with supernova blast waves, while some could be UV-irradiated surfaces of molecular clouds.   Continuation of the CO {\it J}=3--2 survey, as well as follow-up studies of high $R_{\mbox{3--2/1--0}}$ features detected, will reveal the ubiquity and origin of shocked molecular gas in the CMZ.

\bigskip

We thank all members of the ASTE team for excellent performance of the telescope and every support in observations.   This study was financially supported by the MEXT Grant-in-Aid for Scientific Research on Priority Areas No.\ 15071202.  Observations with ASTE were in part carried out remotely from Japan by using NTT's GEMnet2 and its partner R\&E (Research and Education) networks, which are based on AccessNova collaboration of University of Chile, NTT Laboratories, and National Astronomical Observatory of Japan.

\newpage


\end{document}